# Analytical Framework Relating Capacitance-Frequency Measurements and Hysteresis in Perovskite Solar Cells

Manjeet Kumar[1*], Changzeng Ding[1,2], Mathias Nyman[1], Yongqi Sun[2], Chang-Qi Ma[2], Ronald Österbacka[1,2], Oskar J. Sandberg[1*]

1. Physics, Faculty of Science and Engineering, Åbo Akademi University, 20500 Turku, Finland.

2. i- Lab, Suzhou Institute of Nano-Tec and Nano-Bionics, Chinese Academy of Sciences (CAS), 398 Ruoshui Road, SEID, SIP, Suzhou 215123, P. R. China.

E-mail: Manjeet.Kumar@abo.fi, Oskar.Sandberg@abo.fi

## Abstract

Mobile ion migration strongly influences the electrical response and current–voltage hysteresis of perovskite solar cells. However, quantitative determination of ionic transport parameters typically requires transient measurements, extensive numerical simulations, and/or fitting procedures involving multiple uncertain parameters. Here, we present an analytical model that establishes a direct relationship between capacitance–frequency (*C–f)* characteristics and ionic transport properties in perovskite solar cells, while accounting for interlayer effects. The model describes the transition from the ion-screened depletion capacitance at low frequencies to the geometric capacitance at high frequencies and demonstrates that the characteristic transition frequency provides a direct measure of the dielectric relaxation time for ions. Furthermore, the model establishes a quantitative connection between *C–f* characteristics and scan-rate-dependent current–voltage hysteresis, enabling prediction of the scan-rate window in which maximum hysteresis occurs. Drift–diffusion simulations show excellent agreement with the analytical predictions over wide ranges of ion density, ion mobility, transport-layer mobility and doping concentration. Experimental measurements confirm the predicted correlation between capacitance and hysteresis. By providing a direct, physically transparent, and computationally efficient route for extracting ionic transport parameters from routine impedance measurements, this work offers a practical tool for quantifying ion migration and understanding dynamic phenomena in perovskite solar cells.

## Introduction

Metal-halide perovskites have emerged as one of the most important semiconductor materials for next-generation photovoltaics owing to their exceptional optoelectronic properties, low-cost processing, and rapidly increasing power-conversion efficiencies. Since the first demonstration of perovskite-based solar cells, device efficiencies have increased from below 4% to 28%, [1] placing perovskite solar cells (PSCs) among the most promising photovoltaic technologies for future commercialization [2–4]. In addition to their excellent electronic properties, however, metal-halide perovskites exhibit a unique mixed ionic–electronic character, where both electronic charge carriers and mobile ionic defects contribute to the electrical response of the device [5–7].

Unlike in conventional semiconductors, mobile ionic species in perovskite materials can redistribute under an applied electric field and dynamically modify the internal electrostatic potential landscape. This ionic redistribution influences the potential landscape and thereby charge-carrier transport, interfacial charge accumulation, and recombination processes, resulting in a variety of time-dependent phenomena including current–voltage hysteresis, transient responses, dielectric relaxation, and operational instability [8–12]. Ion migration is a fundamental transport process governing the dynamic behavior of perovskite based devices across a broad range of timescales [10,13,14].

Among the various manifestations of ionic transport, current–voltage *(J–V)* hysteresis remains one of the most extensively studied phenomena in PSCs. Hysteresis arises because ionic redistribution occurs on timescales comparable to those of the electrical measurements, causing the internal electric field to lag behind changes in the externally applied bias [7,8,14,15]. Numerous experimental and theoretical studies have shown that the magnitude and position of hysteresis strongly depend on ion migration, measurement conditions, and device architecture [17,11,18,19]. At the same time, ion migration also affects capacitance measurements by inducing space-charge formation that become particularly pronounced at low frequencies [20–23]. Consequently, both frequency-dependent capacitance and hysteresis can be viewed as different manifestations of the same underlying ionic dynamics.

Accurate determination of ionic transport parameters, particularly ion density and ion mobility, is therefore essential for understanding device operation, predicting dynamic behavior, and improving long-term stability. Several approaches have been developed to characterize ionic transport in perovskite semiconductors, including transient ion-drift measurements, transient photocurrent and photovoltage techniques, charge extraction methods, impedance spectroscopy, capacitance spectroscopy, and numerical drift–diffusion simulations [24,25,18,26]. Among these methods, electrochemical impedance spectroscopy and capacitance-frequency (*C–f*) measurements have become particularly attractive because they are experimentally accessible and provide direct information about relaxation processes occurring over a wide range of timescales [11,20,27,28].

Despite considerable progress, reliable extraction of ionic transport parameters remains challenging. Existing approaches often require equivalent-circuit fitting procedures, extensive numerical simulations, or interpretation of transient responses using multiple assumptions regarding dielectric properties, interfacial charge accumulation, recombination pathways, and transport mechanisms [29–31,17,32,33]. As a consequence, reported ion mobility values frequently vary by several orders of magnitude even for similar perovskite compositions [13,21,34]. Recent studies have emphasized that interpretation of frequency-domain measurements requires careful separation of ionic, electronic, and recombination-related contributions, further complicating the quantitative extraction of ionic transport parameters [19,20]. Although *C–f* measurements are widely used to probe ionic dynamics, a simple analytical relationship directly connecting measurable capacitance transitions to ion transport is still lacking.

In this work, we develop an analytical dielectric-relaxation framework that establishes a direct relationship between capacitance–frequency characteristics and ionic transport properties in perovskite solar cells. The model describes the transition from ion-screened depletion capacitance at low frequencies to geometric capacitance at high frequencies and demonstrates that the characteristic transition frequency provides a direct measure of the dielectric relaxation time for ions, from which the ion mobility can be estimated when the ion density is known. Self-consistent drift–diffusion simulations are used to validate the analytical predictions over broad ranges of ion density and mobility in the perovskite layer, and transport-layer conductivity. Crucially, we demonstrate that the same dielectric-relaxation process governing frequency-dependent capacitance also determines the scan-rate dependence of *J–V* hysteresis, enabling prediction of the scan-rate window corresponding to maximum hysteresis. Finally, temperature-dependent capacitance and hysteresis measurements on NiOx/SAM/perovskite/C60 devices provide experimental validation of the proposed framework. The resulting model offers an analytic, physically transparent, and experimentally accessible route for extracting ionic transport parameters and understanding ion-induced dynamic phenomena in perovskite solar cells.

## Theoretical framework of the capacitance

The frequency-dependent response of perovskite solar cells is fundamentally governed by the coexistence of fast electronic transport and comparatively slow ionic migration. Electronic carriers typically respond to electrical perturbations on timescales ranging from picoseconds to microseconds and therefore rapidly establish local electrostatic equilibrium. In contrast, mobile ionic defects redistribute much more slowly [7,35] and may require milliseconds to seconds to reach equilibrium, depending on ion mobility, defect concentration, and electric-field strength.

This separation of timescales introduces a time-dependent electrostatic screening mechanism that strongly modifies the frequency-dependent electrical response of the device, containing direct information of the ionic transport properties. Establishing a quantitative relationship between the frequency-dependent behavior of the associated capacitance and ionic transport forms the basis of the present analytical framework.

An analytical relationship between the total device capacitance and the frequency can be derived based on regional equivalent circuit considerations. We consider a model p-i-n type device with the structure anode | hole-transport layer (HTL) | perovskite active layer (PAL) | electron-transport layer (ETL) | cathode structure. In general, the current flowing through any region $k$ within the device (HTL/PAL/ETL) is composed of the sum of the conduction and displacement currents, spatially averaged across the region. As shown in Supplementary note S1, for small ac voltages, the corresponding conduction and displacement current component translate into a resistor ($R_k$) and capacitor ($C_k$) acting in parallel, with the resulting equivalent impedance $Z_k$ of region $k$ given by $\frac{1}{Z_k} = \frac{1}{R_k} + i\omega C_k$. Here, $R_k = d_k/\sigma_k$ and $C_k = \varepsilon_k/d_k$ represent the equivalent resistance and geometric capacitance of region $k$, being of width $d_k$, conductivity $\sigma_k$, and permittivity $\varepsilon_k$; while $\omega = 2\pi f$ is the angular frequency. Since the total (ac) current flowing through the different regions must be the same, the total impedance is given by the sum of the individual impedances of every region $Z = \sum_k Z_k$. Finally, the device capacitance is obtained from

$$C = \frac{1}{2\pi f} \mathrm{Im}\left(\frac{1}{Z}\right) \tag{1}$$

where $Z$ depends on the frequency $f$ via $\omega$.

For the PAL, we consider the case with mobile anions and an equal number of fixed cations which are uniformly distributed throughout the active layer, with thickness $d_{\mathrm{PAL}}$ and permittivity $\varepsilon_{\mathrm{PAL}}$. Then, provided that the ion density is large enough, the active layer can be divided into three separate regions: a depleted space charge region (SCR) of width $w$ dominated by the space charge of fixed ions (i.e., depleted of mobile ions); a quasi-neutral region (QNR) with uniform density of mobile ions; and a narrow charge accumulation region (CAR) of mobile ions [36]. Furthermore, we assume the conductivity within the depletion region to be negligibly small ($\sigma_{\mathrm{SCR}} \approx 0$). Subsequently, the impedance of the SCR reduces to $Z_{\mathrm{SCR}} = (i\omega C_{\mathrm{SCR}})^{-1}$, with $C_{\mathrm{SCR}} = \varepsilon_{\mathrm{PAL}}/w$. On the other hand, owing to its relatively high ion conductivity and narrow width ($Z_{\mathrm{CAR}} \ll Z_{\mathrm{SCR}}, Z_{\mathrm{QNR}}$), the effect of the narrow ionic accumulation region is omitted. Finally, the resistance, capacitance, and resultant impedance of the neutral region is denoted by $R_{\text{QNR}}$, $C_{\text{QNR}}$, and $Z_{\text{QNR}}$, respectively. As a result, the active layer can be approximated as a capacitor ($C_{\text{SCR}}$) in series with a resistor ($R_{\text{QNR}}$) and capacitance ($C_{\text{QNR}}$) connected in parallel; hence,

$$Z_{\text{PAL}} = \frac{1}{i\omega C_{\text{SCR}}} + \frac{R_{\text{QNR}}}{1+i\omega\tau_{\sigma}} \tag{2}$$

where $\tau_{\sigma} = R_{\text{QNR}}C_{\text{QNR}}$ corresponds to the dielectric relaxation time of ions within the perovskite active layer and is given by [37]

$$\tau_{\sigma} = \frac{\varepsilon_{\text{PAL}}}{qN_{\text{ion}}\mu_{\text{ion}}} \tag{3}$$

assuming the conductivity within the neutral region to be dominated by mobile ions, $\sigma_{\text{QNR}} = qN_{\text{ion}}\mu_{\text{ion}}$, with ion concentration $N_{\text{ion}}$ and ion mobility $\mu_{\text{ion}}$, where $q$ is the elementary charge. Apart from the active layer, a perovskite solar cell device generally also constitutes an HTL and an ETL, each layer having its own impedance. Since these are connected in series, the total impedance of the perovskite solar cell reads $Z = Z_{\text{HTL}} + Z_{\text{PAL}} + Z_{\text{ETL}}$, where $Z_{\text{HTL}}$ ($Z_{\text{ETL}}$) is the corresponding impedances of the HTL (ETL). A schematic picture is shown in Fig 1. The measured *C*–*f* response therefore depends not only on ionic redistribution within the perovskite but also on the properties of the transport layer, including its thickness, dielectric permittivity, and conductivity. Depending on the properties of the interlayers, however, different scenarios can be obtained.

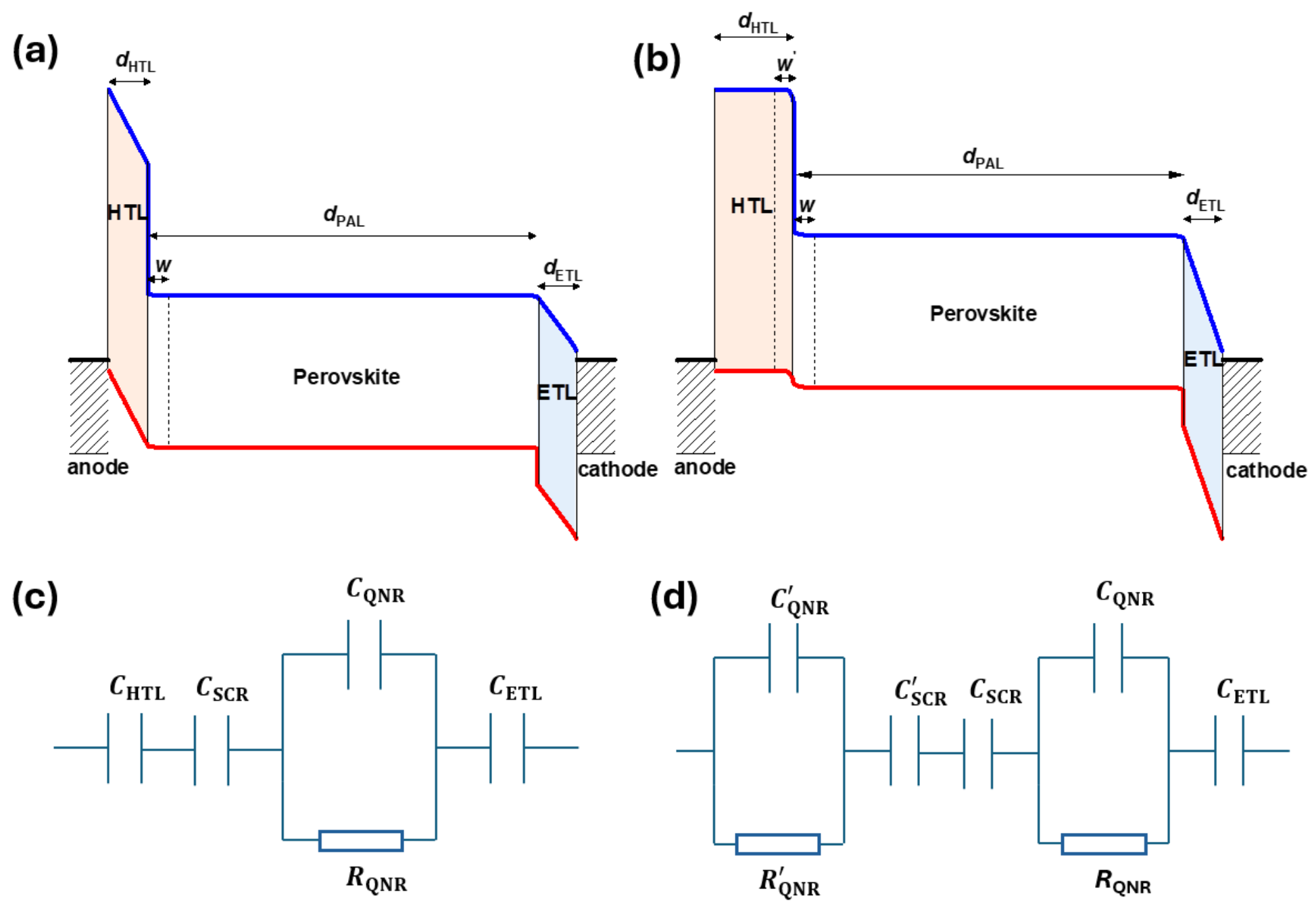

**Fig. 1.** Schematic energy-level diagrams of p-i-n-type perovskite solar cells with a perovskite active layer (PAL) containing a mobile-ion density of $N_{\text{ion}} = 10^{18}$ cm$^{-3}$: **(a)** a device with a thin, undoped interlayer and **(b)** a device with a thick HTL doped to $N_{\text{A}} = 10^{18}$ cm$^{-3}$. The conduction- and valence-band edges are indicated by blue and red solid lines, respectively. **(c, d)** Corresponding equivalent-circuit models for the device configurations shown in panels (a) and (b), respectively.

***The case of undoped interlayers***

We first consider the case of undoped interlayers, assuming the conductivities of both ETL and HTL to be negligibly small (at low voltages). The situation is illustrated in Fig 1(a) and (c). In this limit, the interlayer impedances can both be approximated as $Z_{\mathrm{HTL}} = (i\omega C_{\mathrm{HTL}})^{-1}$ and $Z_{\mathrm{ETL}} = (i\omega C_{\mathrm{ETL}})^{-1}$, with $C_{\mathrm{HTL}} = \varepsilon_{\mathrm{HTL}}/d_{\mathrm{HTL}}$ and $C_{\mathrm{ETL}} = \varepsilon_{\mathrm{ETL}}/d_{\mathrm{ETL}}$ being the capacitance of the HTL and ETL, respectively. Here, $\varepsilon_{\mathrm{HTL}}$ ($\varepsilon_{\mathrm{ETL}}$) is the permittivity and $d_{\mathrm{HTL}}$ ($d_{\mathrm{ETL}}$) is the thickness of the HTL (ETL). The frequency-dependent capacitance is obtained as

$$C = C_{\mathrm{geo}} + \frac{C_{\mathrm{depl}} - C_{\mathrm{geo}}}{1 + \left(2\pi f \tau_{\sigma} \frac{C_{\mathrm{depl}}}{C_{\mathrm{geo}}}\right)^2} \tag{4}$$

where

$$C_{\mathrm{geo}} = \left[\frac{d_{\mathrm{HTL}}}{\varepsilon_{\mathrm{HTL}}} + \frac{d_{\mathrm{PAL}}}{\varepsilon_{\mathrm{PAL}}} + \frac{d_{\mathrm{ETL}}}{\varepsilon_{\mathrm{ETL}}}\right]^{-1} \tag{5}$$

$$C_{\mathrm{depl}} = \left[\frac{d_{\mathrm{HTL}}}{\varepsilon_{\mathrm{HTL}}} + \frac{w}{\varepsilon_{\mathrm{PAL}}} + \frac{d_{\mathrm{ETL}}}{\varepsilon_{\mathrm{ETL}}}\right]^{-1} \tag{6}$$

with $C_{\mathrm{geo}}$ being the total geometric capacitance of the system and $C_{\mathrm{depl}}$ representing an effective depletion capacitance.

The analytical expression Eq. (4) predicts two limiting regimes. In the low-frequency limit, where $2\pi f \tau_{\sigma} \ll 1$, mobile ions fully respond to the external perturbation and efficiently screen the electric field within the active layer. Under these conditions, the measured capacitance approaches the depletion capacitance ($C \to C_{\mathrm{depl}}$) given by Eq. (6). $C_{\mathrm{depl}}$ depends on the ionic depletion layer width $w$ set by the ion density. Accounting for the potential drops across the transport layers (acting as dielectrics), an analytical approximation of $w$ can be derived for high ion densities [36], in this limit Eq. (6) approximates as

$$C_{\mathrm{depl}} \approx \frac{\varepsilon_{\mathrm{PAL}}}{\sqrt{\frac{2\varepsilon_{\mathrm{PAL}}[V_{\mathrm{bi}} - V]}{qN_{\mathrm{ion}}} + \left(\frac{\varepsilon_{\mathrm{PAL}} d_{\mathrm{HTL}}}{\varepsilon_{\mathrm{HTL}}} + \frac{\varepsilon_{\mathrm{PAL}} d_{\mathrm{ETL}}}{\varepsilon_{\mathrm{ETL}}}\right)^2}} \tag{7}$$

where $V_{\mathrm{bi}}$ is the built-in voltage across the device and $V$ is the applied dc voltage (see Supplementary note S2).

On the other hand, in the high-frequency limit, corresponding to $2\pi f \tau_{\sigma} \gg 1$, the measured capacitance approaches the total geometric capacitance ($C \to C_{\mathrm{geo}}$) given by Eq. (5). In this limit, owing to their finite conductivity, ions are no longer able to react to the fast changes in voltage leaving them effectively immobile. The device response then becomes predominantly dielectric, with all the different layers effectively acting as insulators at these frequencies. The transition between these two limits is controlled by $\tau_\sigma$ and therefore directly depends on the ionic conductivity in the active layer.

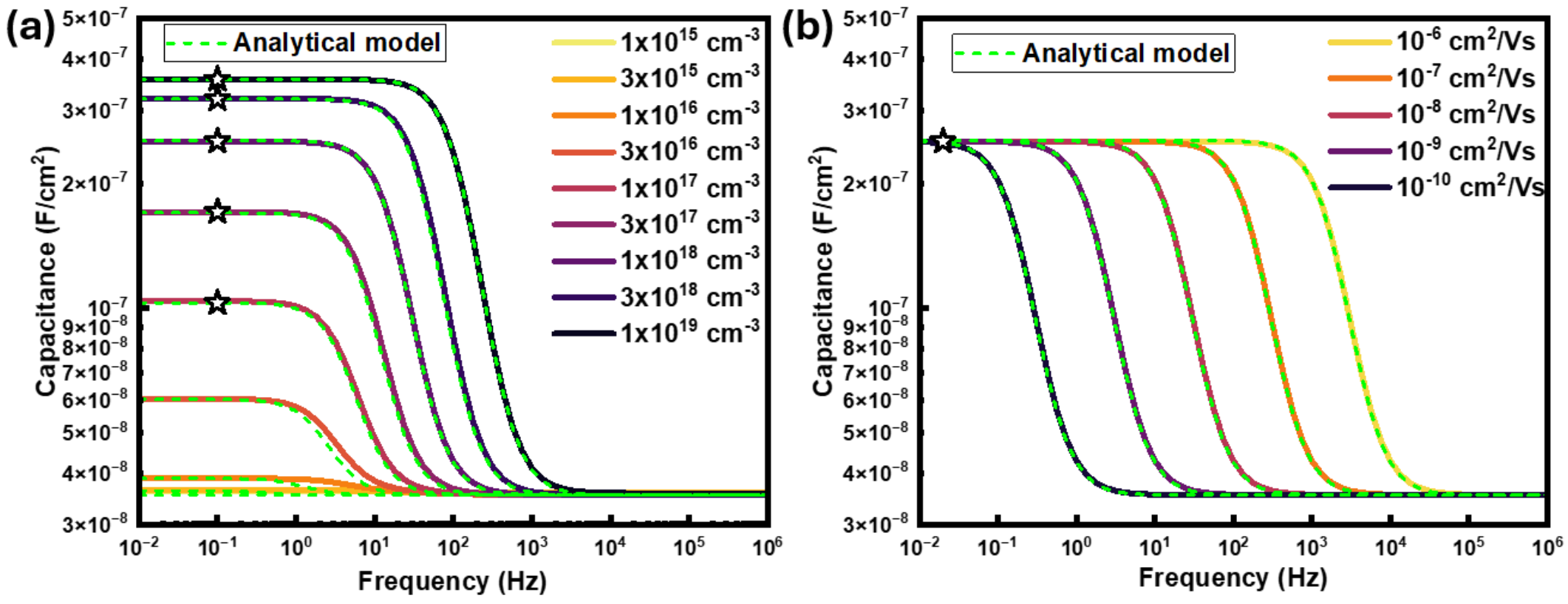

**Fig. 2.** C–f characteristics for a device with a 10 nm ETL and 1 nm HTL, both undoped: (a) effect of varying ion density with ion mobility fixed at $10^{-8}$ $cm^2V^{-1}s^{-1}$, (b) effect of varying ion mobility with ion density fixed at $1x10^{18}$ $cm^{-3}$. The colored solid lines represent results obtained from drift–diffusion simulations, while the green dashed lines indicate the corresponding analytical predictions based on Eq. (4). The star symbols denote the depletion capacitance, $C_{\mathrm{depl}}$, calculated using Eq. (7).

To verify the analytical framework, drift–diffusion simulations were performed for a device structure consisting of a 10 nm ETL and a 1 nm HTL, with both layers assumed to be undoped. Figure 2(a) shows the *C-f* response for different ion densities while maintaining a fixed ion mobility of $10^{-8}$ $\mathrm{cm^2\ V^{-1}\ s^{-1}}$. As the ion density increases, the capacitance transition shifts systematically toward higher frequencies, consistent with the dielectric-relaxation relationship $\tau_\sigma \propto 1/N_{\mathrm{ion}}$. A higher ion density therefore accelerates electrostatic equilibration and enables the ionic distribution to respond to increasingly rapid perturbations. All capacitance spectra exhibit a low-frequency plateau associated with ionic redistribution and depletion-region charging, followed by a transition toward the geometric capacitance at high frequencies where ionic motion can no longer follow the applied field. The analytical expression Eq. (4) accurately reproduces the drift–diffusion simulations across a wide range of ion densities, provided that the ion concentration is sufficiently high to establish effective electrostatic screening and a well-defined depletion region. Under these conditions, the depletion approximation captures the dominant physics governing the capacitance response, with Eq. (7) providing a good approximation for $C_{\mathrm{depl}}$ for the higher ion densities (as indicated by star symbols). At very low ion densities, screening becomes weaker and small deviations from the analytical prediction emerge due to the reduced validity of the depletion approximation. In this limit, $w \rightarrow d_{\mathrm{PAL}}$, with $C_{\mathrm{depl}} \rightarrow C_{\mathrm{geo}}$.

Figure 2(b) examines the influence of ion mobility while maintaining a fixed ion density of $4 \times 10^{17}\ \mathrm{cm^{-3}}$. Since ion mobility directly controls the rate of ionic redistribution, the analytical model predicts $\tau_\sigma \propto 1/\mu_{\mathrm{ion}}$, implying that increasing mobility decreases the dielectric-relaxation time and shifts the capacitance transition toward higher frequencies. The drift–diffusion simulations confirm this prediction over the investigated mobility range from $10^{-10}$ to $10^{-6}\ \mathrm{cm^2\,V^{-1}\,s^{-1}}$. While the transition frequency changes by several orders of magnitude, the capacitance magnitude and overall spectral shape remain nearly unchanged, indicating that ion mobility primarily governs the timescale of ionic redistribution rather than the electrostatic capacitance itself. The analytical model reproduces both the transition frequency and the frequency dependence of capacitance throughout the entire mobility range, demonstrating the robustness of the dielectric-relaxation description.

An important application of this model is that the dielectric relaxation time of ions can be extracted from *C-f* measurements. From Eq. (4) it follows that the inflection point, defining the transition between the low- and high-frequency limits in the *C-f* curve, is given by the characteristic frequency

$$f^* = \frac{C_{\mathrm{geo}}}{2\pi\tau_\sigma C_{\mathrm{depl}}} \tag{8}$$

from which the ionic dielectric relaxation time $\tau_\sigma$ can be directly estimated. This further allows for the ion mobility to be estimated, once the ion density is known (e.g. using Eq. (7)), via $\mu_{\mathrm{ion}} = \frac{\varepsilon_{\mathrm{PAL}}\varepsilon_0}{q\tau_\sigma N_{\mathrm{ion}}}$. This relation constitutes a central result of the present work. Rather than relying on extensive numerical fitting, or multiple uncertain fitting parameters, ion mobility can be obtained directly from experimentally accessible frequency-dependent capacitance measurements. The methodology requires only determination of the geometric capacitance, depletion capacitance, transition frequency, and an independent estimate of ion density obtained, for example, from capacitance–voltage measurements or electrostatic analysis.

***The case of doped interlayers***

The analytical approximation presented above assumes that the transport layers are sufficiently thin to either act like undoped layers or their capacitance is much larger than the capacitance of the ionic space charge region in the perovskite. However, many device architectures contain relatively thick and doped transport layers. Under these conditions, the measured *C–f* response will also depend on the (electronic) conductivity of the transport layers. To investigate this regime, the above case was extended to account for the case of a p-doped HTL while keeping the ETL undoped. This situation is depicted in Fig. 1(b) and (d).

Provided that the doping concentration is high enough, the p-doped HTL is divided into a field-free quasi-neutral region, of width $d_{\mathrm{HTL}} - w'$ and conductivity $\sigma'_{\mathrm{QNR}} = qN_{\mathrm{dop}}\mu_{\mathrm{HTL}}$, and a hole-depleted space charge region of width $w'$. $N_{\mathrm{dop}}$ and $\mu_{\mathrm{HTL}}$ represent the doping concentration and hole mobility inside the transport layer. In analogy with Eq. (2), the impedance of the HTL is given by $Z_{\mathrm{HTL}} = \frac{1}{i\omega C'_{\mathrm{SCR}}} + \frac{R'_{\mathrm{QNR}}}{1+i\omega\tau'_{\sigma}}$, where $R'_{\mathrm{QNR}} = [d_{\mathrm{HTL}} - w']/\sigma'_{\mathrm{QNR}}$ and $C'_{\mathrm{SCR}} = \varepsilon_{\mathrm{HTL}}/w'$ is the resistance of the HTL neutral region and capacitance of the HTL depletion region, respectively; while

$$\tau'_{\sigma} = \frac{\varepsilon_{\mathrm{HTL}}}{qN_{\mathrm{dop}}\mu_{\mathrm{interlayer}}} \tag{9}$$

is the dielectric relaxation time of holes within the transport layer.

Under these conditions, the resulting capacitance response of the multilayer device can be obtained as

$$C = C'_{\mathrm{depl}} \frac{1+\frac{4\pi^2 f^2 \tau_{\sigma}^2 C'_{\mathrm{depl}}}{C'_{\mathrm{geo}}}\left[1+\frac{4\pi^2 f^2 \tau'^2_{\sigma} C'_{\mathrm{geo}}}{C_{\mathrm{geo}}}\right]}{1+4\pi^2 f^2 \tau_{\sigma}^2 \left(\frac{C'_{\mathrm{depl}}}{C'_{\mathrm{geo}}}\right)^2 \left[1+4\pi^2 f^2 \tau'^2_{\sigma}\left(\frac{C'_{\mathrm{geo}}}{C_{\mathrm{geo}}}\right)^2\right]} \tag{10}$$

where

$$C'_{\mathrm{geo}} = \left[\frac{w'}{\varepsilon_{\mathrm{HTL}}} + \frac{d_{\mathrm{PAL}}}{\varepsilon_{\mathrm{PAL}}} + \frac{d_{\mathrm{ETL}}}{\varepsilon_{\mathrm{ETL}}}\right]^{-1} \tag{11}$$

$$C'_{\mathrm{depl}} = \left[\frac{w'}{\varepsilon_{\mathrm{HTL}}} + \frac{w}{\varepsilon_{\mathrm{PAL}}} + \frac{d_{\mathrm{ETL}}}{\varepsilon_{\mathrm{ETL}}}\right]^{-1} \tag{12}$$

Here, $C'_{\mathrm{geo}}$ ($C'_{\mathrm{depl}}$) is the equivalent capacitance of the HTL depletion capacitance and the ETL geometric capacitance in series with the geometric (ionic depletion) capacitance of the active layer, while $C_{\mathrm{geo}}$ is the total geometric capacitance [Eq. (5)] as before.

Eq. (10) predicts the coexistence of two distinct relaxation processes originating from ionic redistribution and electronic transport-layer equilibration. Under experimentally relevant conditions where the dielectric relaxation of ions is substantially slower than transport-layer equilibration ($\tau_{\sigma} \gg \tau'_{\sigma}$), the capacitance response separates naturally into two characteristic regimes. At low frequencies, the transport layers respond rapidly, and the capacitance remains dominated by ionic redistribution. For low frequencies, Eq. (10) can therefore be approximated as

$$C = C'_{\text{geo}} + \frac{C'_{\text{depl}} - C'_{\text{geo}}}{1 + \left(2\pi f \tau_{\sigma} \frac{C'_{\text{depl}}}{C'_{\text{geo}}}\right)^2} \tag{13}$$

demonstrating that the low-frequency transition remains governed by ionic dielectric relaxation. However, in this case the low-frequency plateau is governed by a combined depletion layer constituting the (ionic) depletion layer within the active layer and the (electronic) depletion layer in the HTL. The situation is analogous to that of a classical p-n heterojunction. Hence, after accounting for the potential drops across the two depletion layers and the ETL, we find

$$C'_{\text{depl}} \approx \frac{\varepsilon_{\text{PAL}}}{\sqrt{\frac{2\varepsilon_{\text{PAL}}[V_{\text{bi}} - V]}{qN_{\text{ion}}}\left(1 + \frac{\varepsilon_{\text{PAL}} N_{\text{ion}}}{\varepsilon_{\text{HTL}} N_{\text{dop}}} +\right) + \left(\frac{\varepsilon_{\text{PAL}} d_{\text{ETL}}}{\varepsilon_{\text{ETL}}}\right)^2}} \tag{14}$$

provided that $N_{\text{ion}}$ and $N_{\text{dop}}$ are both high enough. In this limit, we expect the capacitance to depend on both $N_{\text{ion}}$ and $N_{\text{dop}}$.

At sufficiently high frequencies, ionic motion becomes frozen, and the capacitance reaches an intermediate-frequency plateau, $C \to C'_{\text{geo}}$. In this regime, the active layer effectively acts like a dielectric. The situation is otherwise identical to the case in Eq. (7), but with HTL and active layer swapping roles; hence,

$$C'_{\text{geo}} \approx \frac{\varepsilon_{\text{HTL}}}{\sqrt{\frac{2\varepsilon_{\text{HTL}}[V_{\text{bi}} - V]}{qN_{\text{dop}}} + \left(\frac{\varepsilon_{\text{HTL}} d_{\text{PAL}}}{\varepsilon_{\text{PAL}}} + \frac{\varepsilon_{\text{HTL}} d_{\text{ETL}}}{\varepsilon_{\text{ETL}}}\right)^2}} \tag{15}$$

provided $N_{\text{dop}}$ is high enough. In accordance with Eq. (15), we expect the intermediate-frequency capacitance plateau to depend on the doping concentration in the HTL, but to be largely independent of ion concentration within the active layer in this limit.

For even higher frequencies, the remaining capacitance variation originates solely from transport-layer relaxation, and Eq. (10) simplifies as

$$C = C_{\text{geo}} + \frac{C'_{\text{geo}} - C_{\text{geo}}}{1 + \left(2\pi f \tau'_{\sigma} \frac{C'_{\text{geo}}}{C_{\text{geo}}}\right)^2} \tag{16}$$

For high enough frequencies the charge carrier motion in the HTL is eventually frozen, as the total geometric capacitance of the system is reached ($C \to C_{\text{geo}}$). The analytical model therefore predicts that multilayer devices should exhibit two separated capacitance transitions corresponding to ionic and transport-layer relaxation processes.

To validate the extended analytical framework, numerical drift–diffusion simulations were performed for the multilayer structure containing a 10 nm undoped ETL and a 100 nm p-doped HTL. Unless otherwise specified, the HTL parameters were fixed at a doping concentration of $1 \times 10^{18}$ cm$^{-3}$, hole mobility of $1.5 \times 10^{-4}$ cm$^2$V$^{-1}$s$^{-1}$ and the active layer contained mobile ions with an ion density of $1 \times 10^{18}$ cm$^{-3}$ and an ion mobility of $1 \times 10^{-8}$ cm$^2$ V$^{-1}$ s$^{-1}$. The simulated *C-f* characteristics were then directly compared with the analytical model Eq. (10).

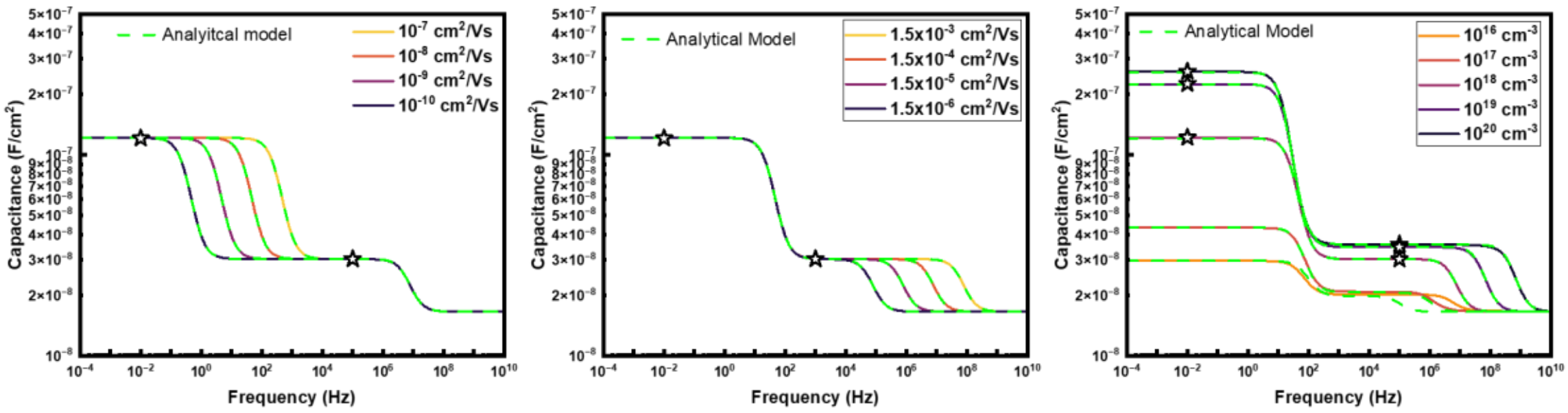

**Fig. 3.** C–f characteristics for a device with a 10 nm undoped ETL and a 100 nm HTL. Unless otherwise specified, the HTL parameters were fixed at a doping concentration of $1 \times 10^{18}$ cm$^{-3}$, hole mobility of $1.5 \times 10^{-4}$ cm$^2$ V$^{-1}$s$^{-1}$ and the active layer contained mobile ions with an ion density of $1 \times 10^{18}$ cm$^{-3}$ and an ion mobility of $1 \times 10^{-8}$ cm$^2$ V$^{-1}$ s$^{-1}$. (a) Ion mobility is varied; (b) hole mobility is varied; (c) HTL doping concentration is varied. The colored solid lines represent drift–diffusion simulation results, while green dashed lines depict the corresponding analytical predictions based on Eq. (10). While the star-shaped symbols represent the extracted values of $C'_{\mathrm{depl}}$ and $C'_{\mathrm{geo}}$, determined from Eqs. (14) and (15), respectively.

Figure 3(a,b) compares the effects of ion mobility and HTL hole mobility while all other parameters remain constant (see Supplementary Table S1). Increasing ion mobility shifts the low-frequency transition towards higher frequencies by reducing the ionic relaxation time, while the high-frequency transition remains nearly unaffected. In contrast, decreasing the HTL hole mobility reduces the HTL conductivity, slows interlayer charge equilibration, and shifts the high-frequency transition toward lower frequencies, while leaving the low-frequency ionic transition nearly unchanged. The excellent agreement between analytical Eq. (10) and drift–diffusion simulation results confirm that the two transitions originate from ionic dielectric relaxation and transport-layer charge equilibration, respectively. The star symbols in all panels represent the values of $C'_{\mathrm{depl}}$ and $C'_{\mathrm{geo}}$, calculated using Eqs. (14) and (15), respectively, showing good agreement for high ion and doping densities, as expected.

Figure 3(c) explores the effect of HTL doping concentration while all other parameters are constant. Increasing the doping concentration alters electrostatic screening within the transport layer and modifies the coupling between the active layer and the HTL. As a result, both the capacitance magnitude and the transition characteristics evolve systematically with doping concentration. Enhanced doping further sharpens the capacitance transitions and increases the separation between the different dynamic regimes. The analytical model Eq. (10) successfully captures these trends and remains in quantitative agreement with the numerical simulations over several orders of magnitude in doping concentration. These results demonstrate that the proposed framework retains its predictive accuracy even for realistic multilayer device architectures incorporating thick transport layers.

The overall excellent agreement between the analytical and numerical results demonstrates that the dielectric relaxation framework captures the key physical processes responsible for frequency-dependent capacitance also for the case of thick interlayer devices. Importantly, the validity of the model extends beyond variations in ionic properties and remains robust for changes in transport-layer carrier mobility and doping concentration, demonstrating the generality and robustness of the analytical treatment.

**Relating ionic relaxation to hysteresis**

*J–V* hysteresis in perovskite solar cells is mainly associated with the ion migration and redistribution of mobile ionic defects under an applied electric field. Since both low frequency capacitance response and *J–V* hysteresis originates from the same underlying ionic dynamics, *C–f* measurements provide a direct means of characterizing the processes responsible for hysteretic behavior. Although capacitance spectroscopy and *J–V* scans are performed under different experimental conditions, both are affected by the same fundamental physical process: the delayed response of mobile ions. In *C–f* measurements, this process is observed in the frequency domain under small periodic perturbations, whereas during *J–V* scans the applied voltage changes continuously with time. In both cases, the finite ionic mobility prevents the internal electrostatic potential from instantaneously following the externally applied bias. Consequently, the ionic system exhibits a characteristic response time that governs its departure from and return to equilibrium.

Establishing a correlation between *C–f* measurements and *J–V* hysteresis is therefore valuable because the ionic relaxation time extracted from capacitance spectroscopy can be directly linked to the timescale over which non-equilibrium ionic distributions develop during voltage scanning. This approach identifies the characteristic scan timescale at which ionic redistribution is most strongly delayed relative to the applied bias.

During a *J-V* scan the applied voltage depends on the time $t$. The device is initially held at steady state at a pre-bias voltage $V_{\text{pre}}$. At $t = 0$, a reverse scan $V(t) = V_{\text{pre}} - At$ is applied until $t = t_{\text{pulse}}$, where $A = V_{\text{span}}/t_{\text{pulse}}$ is the scan rate, $V_{\text{span}} = |V_{max} - V_{min}|$ is the voltage span, and $t_{\text{pulse}}$ is the pulse duration. At $t = t_{\text{pulse}}$, the scanning direction is reversed and a forward scan $V(t) = V_{\text{pre}} + A\left[t - 2t_{\text{pulse}}\right]$ is applied. Finally, at $t = 2t_{\text{pulse}}$ the applied voltage has returned to its starting value $V\left(2t_{\text{pulse}}\right) = V_{\text{pre}}$, and the *J-V* scan is terminated. The degree of hysteresis depends on how much the current, when evaluated at the same applied voltage, differs between forward and reverse scan. The scan direction dependence is generally attributed to differences in the potential landscape due to ion-induced field screening. Since the electric field $E_{\mathrm{QNR}}(t)$ in the quasi-neutral region of the active layer is a direct measure of this field screening, the degree of hysteresis depends on how much $E_{\mathrm{QNR}}$ at the end ($t = 2t_{\mathrm{pulse}}$) of the scan differs from the starting point ($t = 0$). In other words, we expect hysteresis to be maximal near the point where $\Delta E_{\mathrm{QNR}} = |E_{\mathrm{QNR}}(2t_{\mathrm{pulse}}) - E_{\mathrm{QNR}}(0)|$ is at its largest [38].

Provided that the ion density is high enough, the electric field in the neutral region is initially screened (at $t = 0$), and the applied voltage within the perovskite layer drops across the ion depletion region. Applying the voltage pulse at $t > 0$ will disturb the electric field in the device, resulting in a displacement of the mobile ions within the active layer. The subsequent change $\Delta w(t)$ in the depletion region is governed by $\frac{d\Delta w}{dt} = -\mu_{\mathrm{ion}} E_{\mathrm{QNR}}(t)$ [39,40]. For the case of undoped transport layers, after accounting for the potential drop across the different layers and the screening (space charge) effects of ions, we find (see Supplementary note S2)

$$\frac{d\Delta w(t)}{dt} + \frac{[w_{\mathrm{eff}} + \Delta w(t)]^2 - w_{\mathrm{eff}}^2}{2\tau_\sigma d_{\mathrm{eff}}} = -\frac{\mu_{\mathrm{ion}}[V(t) - V(0)]}{d_{\mathrm{eff}}} \quad (17)$$

Here, $d_{\mathrm{eff}} = \varepsilon_{\mathrm{PAL}}/C_{\mathrm{geo}}$ with $C_{\mathrm{geo}}$ given by Eq. (5), while $w_{\mathrm{eff}} = \varepsilon_{\mathrm{PAL}}/C_{\mathrm{depl}}$ is the effective depletion region width at $t = 0$, where $C_{\mathrm{depl}}$ is defined by Eq. (6) being the effective depletion capacitance at $V = V_{\mathrm{pre}}$.

Based on Eq. (17), two extreme cases can directly be identified. For short pulses (fast scans), when $t \ll \tau_\sigma$, the second term on the left-hand side of Eq. (17) becomes negligibly small. In this limit, owing to their finite mobility, the ions are essentially frozen and unable to readjust and screen the electric field within the active layer. As a result, the time-dependent applied voltage component drops uniformly across the entire device, with $E_{\mathrm{QNR}}(t) = [V(t) - V(0)]/d_{\mathrm{eff}}$. Since $V(2t_{\mathrm{pulse}}) = V(0)$, we thus expect hysteresis induced by field screening to be negligible in this limit. For long voltage pulses (fast scans), in turn, when $t \gg \tau_\sigma$, the first term on the left-hand side of Eq. (17) becomes vanishingly small. In this limit, ions have enough time to fully equilibrate and redistribute in accordance with the changing voltage. Subsequently, the electric field within neutral region remains perpetually screened, $E_{\mathrm{QNR}}(t) \to 0$, as the entire applied voltage within the active layer drops across the depletion region, suggesting vanishingly small hysteresis in this extreme.

The above findings suggest negligible hysteresis ($\Delta E_{\mathrm{QNR}} \to 0$) for the extreme cases of weak and strong field screening during the scan. Instead, hysteresis predominantly occurs in the intermediate case when $E_{\mathrm{QNR}}(t)$ has been partially screened. This corresponds to a situation when the pulse is long enough for the second term on the left-hand side of Eq. (17) to become significant, but for $\Delta w(t)$ to still be small not to significantly screen the electric field. Under these conditions, corresponding to $\Delta w(t) \ll w$, solving Eq. (17) yields

$$\Delta E_{\mathrm{QNR}} \approx \frac{V_{\mathrm{span}}}{d_{\mathrm{eff}}} \times \frac{\left[1-\left(1+\frac{t_{\mathrm{pulse}}}{\tau_\sigma}\times\frac{C_{\mathrm{geo}}}{C_{\mathrm{depl}}}\right)\exp\left(-\frac{t_{\mathrm{pulse}}}{\tau_\sigma}\times\frac{C_{\mathrm{geo}}}{C_{\mathrm{depl}}}\right)\right]}{\left(\frac{t_{\mathrm{pulse}}}{\tau_\sigma}\times\frac{C_{\mathrm{geo}}}{C_{\mathrm{depl}}}\right)} \tag{18}$$

as shown in Supplementary note S2. As expected, Eq. (18) reproduces $\Delta E_{\mathrm{QNR}} \to 0$ in the limits of small and large $t_{\mathrm{pulse}}$. Conversely, Eq. (18) has its maximal value at

$$t_{\mathrm{pulse}}|_{\max} \approx 1.79 \times \tau_\sigma \frac{C_{\mathrm{depl}}}{C_{\mathrm{geo}}} \tag{19}$$

yielding the scan duration at which the strongest hysteretic behavior is expected to occur.
Eq. (19) can be used to predict the time-window, and the corresponding scan rate $A|_{\max} = V_{\mathrm{span}}/t_{\mathrm{pulse}}|_{\max}$, at which the system suffers from hysteresis induced by ionic field screening during the voltage sweep. For high enough ion densities, the capacitance does not change considerably with applied voltage (see supplementary Fig. S3). In this case, comparing Eq. (8) and (19) reveals that the predicted time scale of hysteresis is directly linked to the ionic transition frequency $f^*$of the *C-f* curve via $t_{\mathrm{pulse}}|_{\max} \approx 0.285/f^*$. It should be emphasized that the present model does not predict the magnitude of the hysteresis index, as this quantity is additionally influenced by recombination processes, interface properties, transport layers, defect distributions, and measurement history. Thus, the ionic dielectric relaxation time serves as a predictor of the relevant timescale rather than the amplitude of the hysteretic response.
To validate Eq. (19), scan-rate-dependent drift–diffusion simulations were analyzed using the hysteresis index $\mathrm{HI} = (\mathrm{PCE_{rs}} - \mathrm{PCE_{fs}})/\mathrm{PCE_{rs}}$, where $\mathrm{PCE_{rs}}$ and $\mathrm{PCE_{fs}}$ denote the reverse-scan and forward-scan efficiencies, respectively. Figure 3 presents the simulated hysteresis index under different ionic conditions. The colored solid curves correspond to drift–diffusion simulation results, whereas the arrows represent the analytical predictions $A|_{\max} = V_{\mathrm{span}}/t_{\mathrm{pulse}}|_{\max}$ obtained from Eq. (19). The close agreement between the predicted arrow positions and the maxima obtained from the numerical simulations demonstrates that the hysteresis peak position is fundamentally governed by the ionic dielectric relaxation timescale. Figures 4(a) and 4(b) correspond to the thin-interlayer device architecture introduced in Figures 2(a) and 2(b), enabling a direct comparison between the capacitance-derived relaxation times and scan-rate-dependent hysteresis behavior. Figure 4(a) examines the influence of ion density while maintaining a constant ion mobility. Increasing the ion density shifts the hysteresis peak toward higher scan rates because the ionic dielectric relaxation time decreases with increasing ionic concentration ($\tau_\sigma \propto 1/N_{ion}$). A higher ionic density therefore accelerates electrostatic relaxation, requiring faster voltage scans to drive the ionic distribution out of equilibrium. The amplitude of the maximum hysteresis index also depends on the ion density, interlayer properties, and bulk SRH recombination lifetime (see supplementary fig S2). The analytical predictions accurately reproduce the simulated peak positions over several orders of magnitude variation in ion density. Figure 4(b) investigates the effect of ion mobility while keeping the ion density constant. In agreement with Eq. (19) (with $\tau_\sigma \propto 1/\mu_{ion}$), increasing ion mobility reduces the ionic equilibration time and shifts the maximum hysteresis toward higher scan rates. Consequently, faster voltage sweeps are required to prevent the ionic distribution from equilibrating and following the applied bias. The analytically predicted peak positions closely match those obtained from the drift–diffusion simulations, confirming that the same dielectric relaxation process responsible for the frequency-dependent capacitance in Figure 1 also governs the scan-rate dependence of hysteresis.

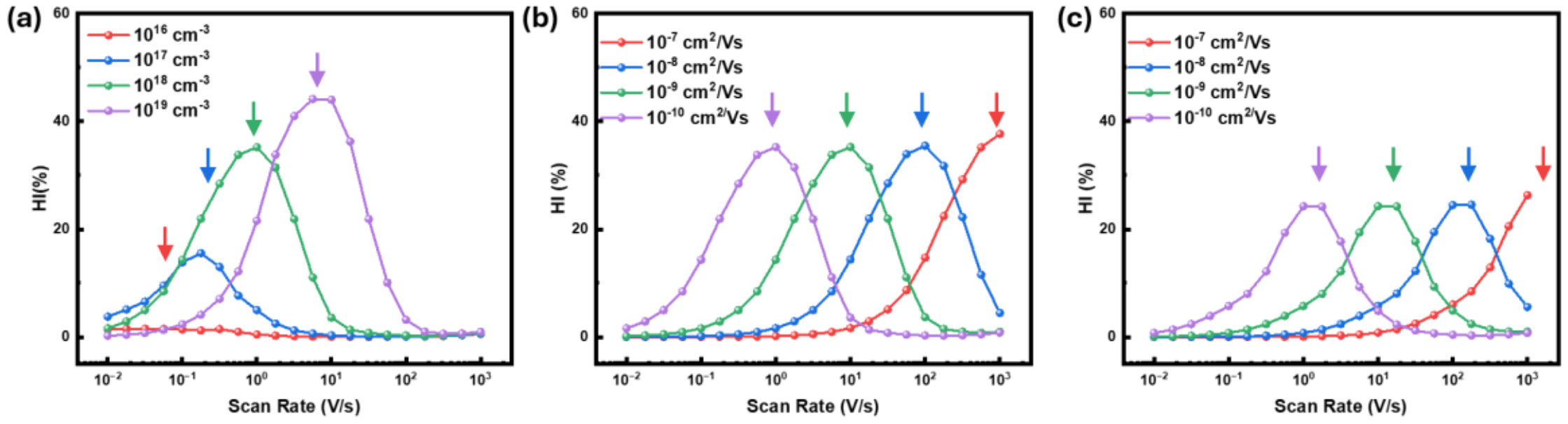

**Fig. 4.** Hysteresis index (HI), defined as HI = ($PCE_{rs}$- $PCE_{fs}$)/$PCE_{rs}$, illustrating the impact of ionic and transport properties under different device configurations. Panels (a) and (b) correspond to a structure comprising a 10 nm ETL and a 1 nm undoped HTL. In panel (a), the ion mobility is fixed $1 \times 10^{-10}$ cm$^2$ V$^{-1}$ s$^{-1}$ while the ion density is varied; in panel (b), the ion density is fixed $1 \times 10^{18}$ cm$^{-3}$ while the ion mobility is varied. Panel (c) represents a device with a 10 nm ETL and a 100 nm HTL doped at $1 \times 10^{18}$ cm$^{-3}$, where ion mobility is varied. The colored solid lines denote results from drift-diffusion simulations, while arrows indicate the analytical predictions, based on Eq. (19), for the scan rate at which the hysteresis index reaches its maximum (i.e., the position of the peak rather than its magnitude).

The analysis was subsequently extended to the multilayer device architecture shown in Figure 3(a). Figure 4(c) presents simulations for a structure containing a 10 nm ETL and a 100 nm HTL with a HTL doping concentration of $1 \times 10^{18}$ cm$^{-3}$. Despite the increased electrostatic complexity introduced by the transport layers, the same fundamental behavior is observed. A pronounced hysteresis maximum appears at an intermediate scan rate, and variations in ion mobility systematically shift the peak position. The analytical predictions continue to accurately capture the simulated maxima, where we effectively replaced $C_{\mathrm{depl}}/C_{\mathrm{geo}}$ with $C'_{\mathrm{depl}}/C'_{\mathrm{geo}}$ in Eq. (19) to account for interlayer depletion. This demonstrates that ionic dielectric relaxation remains the dominant timescale governing hysteresis even in realistic multilayer devices. Although the transport layers influence the detailed shape and magnitude of the hysteresis response, they do not alter the fundamental relationship between ionic relaxation and the scan rate corresponding to maximum hysteresis.

These results establish a unified interpretation of frequency-dependent capacitance and *J–V* hysteresis in perovskite solar cells. The capacitance transition observed in *C-f* measurements represent the frequency-domain manifestation of ionic dielectric relaxation, whereas the hysteresis peak observed in scan-rate-dependent *J–V* measurements represent the corresponding time-domain manifestation of the same process. Consequently, capacitance spectroscopy provides not only a means of extracting ionic transport parameters but also a predictive tool for identifying the scan-rate window in which hysteresis is expected to occur.

**Experimental results**

Temperature-dependent measurements provide a particularly powerful way to validate the proposed dielectric relaxation framework because ionic transport is a thermally activated process. As temperature increases, ion mobility increases, directly altering the timescale of ionic redistribution and, consequently, the ionic dielectric relaxation dynamics within the device. This temperature dependence offers an opportunity to examine whether the capacitance response and *J–V* hysteresis are governed by the same underlying ionic processes. To investigate this relationship experimentally, temperature-dependent *C-f* measurements were performed in the dark, while hysteresis measurements were conducted under blue-light illumination with an intensity of approximately 0.5 sun using devices with the structure ITO/$NiO_x$/SAM/perovskite/$C_{60}$/Au. If both the low-frequency capacitance response and the scan-rate-dependent hysteresis originate from mobile ion redistribution, their characteristic timescales should exhibit consistent temperature dependence. Therefore, changes in temperature are expected to produce corresponding shifts in both the capacitance transition frequency and the hysteresis behavior, providing a direct experimental validation of the common ionic dielectric relaxation mechanism linking these two phenomena.

Figure 5(a) presents the temperature-dependent *C-f* characteristics. As the temperature decreases, the capacitance transition systematically shifts toward lower frequencies, indicating that the ionic redistribution process becomes progressively slower. This behavior is expected because ion migration in perovskite materials is thermally activated; lower temperatures reduce the rate at which mobile ions can respond to changes in the electric field. Consequently, the characteristic dielectric relaxation time increases, and the transition from the low-frequency ion-screened capacitance to the high-frequency geometric capacitance occurs on longer timescales.

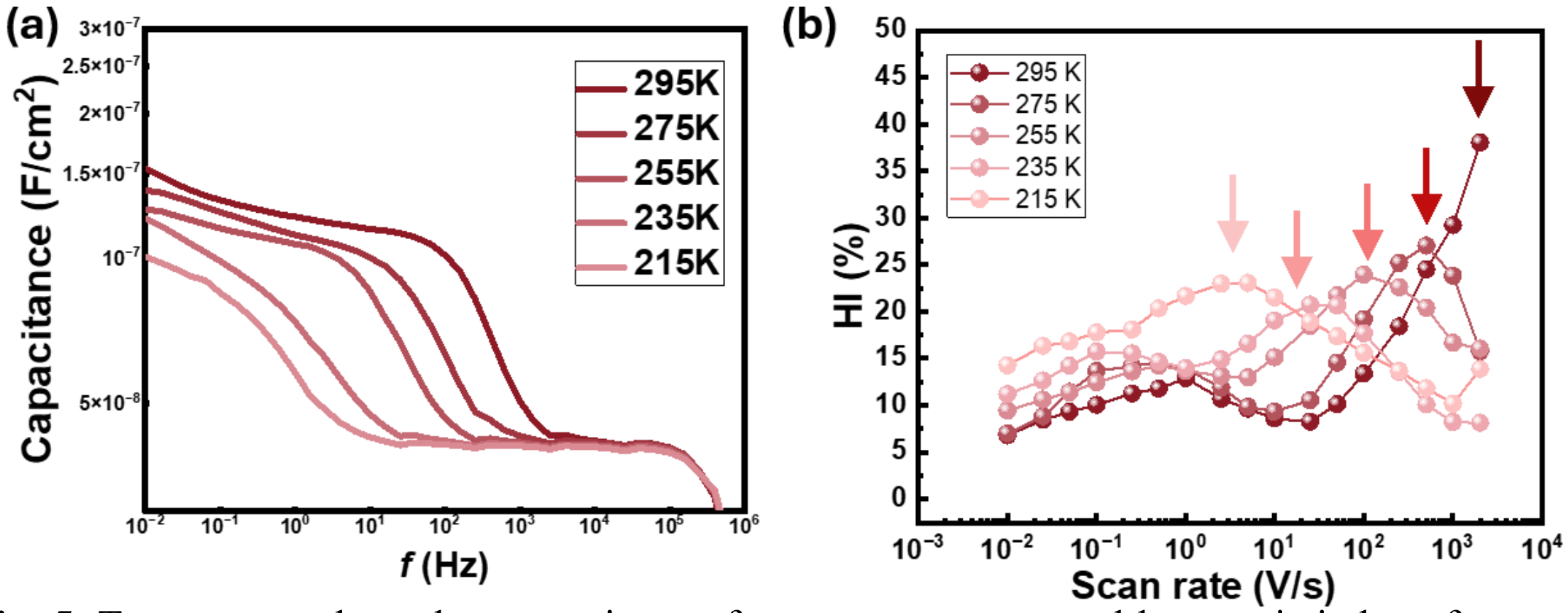

**Fig. 5.** Temperature-dependent capacitance–frequency response and hysteresis index of perovskite solar cells. (a) *C–f* characteristics of the perovskite solar cell measured at different temperatures, showing the temperature-activated capacitive response associated with ionic/electronic processes and interfacial charge accumulation. (b) Corresponding temperature-dependent hysteresis index, extracted from the forward and reverse scan current–voltage characteristics, illustrating the evolution of device hysteresis with temperature. The colored solid lines denote experimental results, while arrows indicate the analytical predictions, based on Eq. (19), for the scan rate at which the hysteresis index reaches its maximum (i.e., the position of the peak rather than its magnitude).

To quantify this behavior, the *C-f* characteristics at each temperature were fitted using the analytical model Eq. (4) (see Supplementary Note S3). The fitting yields the characteristic dielectric relaxation time, which represents the timescale required for ionic charge redistribution within the device. Consistent with the observed shift in the capacitance transition, the extracted relaxation time increases systematically with decreasing temperature, reflecting the slower ionic dynamics at lower temperatures. Since the same ionic redistribution process is expected to be responsible for scan-rate-dependent hysteresis, the scaled dielectric relaxation time $\tau_\sigma C_{\mathrm{depl}}/C_{\mathrm{geo}}$ obtained from the capacitance measurements can be used directly to predict the scan-rate position at which the hysteresis index is expected to reach its maximum.

Figure 5(b) shows the experimentally measured hysteresis index as a function of scan rate for different temperatures. At all temperatures, the hysteresis exhibits a pronounced non-monotonic dependence on scan rate, with a distinct maximum occurring at an intermediate scan rate. As the temperature decreases, the hysteresis maximum shifts systematically toward lower scan rates, indicating that longer measurement times are required for the ionic system to respond.

The arrows in Figure 5(b) indicate the analytically predicted scan-rate positions corresponding to maximum hysteresis, calculated directly from Eq. (19) using the dielectric relaxation times extracted from the *C–f* measurements (see Supplementary Note S3 and Table S2). A very good agreement is observed between the experimentally measured hysteresis maxima and the predicted positions across the entire temperature range. Note that the analytical model is intended to predict the characteristic scan-rate position of the hysteresis maximum rather than the absolute magnitude which may additionally be influenced by factors such as recombination dynamics, interface properties, and measurement history. Nevertheless, the excellent agreement between experiment and prediction demonstrates that the ionic dielectric relaxation time obtained from *C-f* measurements captures the dominant timescale governing hysteresis evolution. These results establish a direct and quantitative connection between capacitance relaxation and hysteresis dynamics, providing strong evidence that ionic dielectric relaxation is the primary mechanism controlling the scan-rate dependence of hysteresis in these devices.

The above findings experimentally substantiate the link between *C-f* and hysteresis. Based on this, a method to experimentally predict hysteresis dynamics can be devised. First, the *C-f* response of the perovskite solar cell is measured, allowing for $C_{\mathrm{geo}}$ and $C_{\mathrm{depl}}$ to be estimated. Equation (4) is then fitted to the measured *C-f* spectrum to obtain the scaled ionic dielectric relaxation time ($\tau_\sigma C_{\mathrm{depl}}/C_{\mathrm{geo}}$) from which $\tau_\sigma$ can be calculated. By inserting the scaled ionic dielectric relaxation time into Eq. (19), $t_{\mathrm{pulse|max}}$ can be obtained. Finally, the scan rate corresponding to the maximum hysteresis index can then be estimated as $V_{\mathrm{span}}/t_{\mathrm{pulse|max}}$.

## Conclusions

In this work, we developed an analytical framework that links the frequency-dependent capacitance response of perovskite solar cells directly to ionic transport properties through ionic dielectric relaxation. The model demonstrates that the transition from low-frequency depletion capacitance to high-frequency geometric capacitance is governed by the characteristic ionic relaxation time, providing a straightforward route for determining ion mobility from experimentally accessible *C–f* measurements when the ion density is known. Unlike conventional approaches that often require complex transient techniques or extensive numerical fitting, the proposed method offers a physically transparent and computationally efficient alternative. The analytical description was validated through self-consistent drift–diffusion simulations over wide ranges of ion density and ion mobility. The framework was further extended to device architectures containing thick, doped transport layers, revealing the coexistence of ionic and transport-layer relaxation processes and accurately reproducing the resulting frequency-dependent capacitance response. The excellent agreement between analytical predictions and numerical simulations demonstrates the robustness and general applicability of the model. Furthermore, we established a direct connection between the frequency-dependent capacitance response and scan-rate-dependent current–voltage hysteresis. The same ionic relaxation time extracted from *C–f* measurements was shown to predict the characteristic scan-rate window at which hysteresis reaches its maximum. Experimental temperature-dependent measurements confirmed this relationship, providing strong evidence that ionic dielectric relaxation is the common physical origin underlying both the frequency-dependent capacitance response and hysteresis dynamics.

Overall, this work provides a unified description of ion-induced dynamic phenomena in perovskite solar cells by linking ionic dielectric relaxation, the frequency-dependent capacitance response, and hysteresis dynamics within a single analytical framework. The proposed model offers a simple and practical approach for extracting ion mobility from routine capacitance–frequency measurements and provides valuable insight into the role of ion migration in device operation and stability. More broadly, the framework establishes capacitance spectroscopy as a powerful tool for investigating ionic transport and predicting dynamic device behavior in perovskite solar cells and related mixed ionic–electronic conductors.

**Acknowledgements**

This work is supported by the Finnish Ministry of Education and Culture through the PREIN/I-DEEP doctoral pilot (VN/3137/2024-OKM-4), the National Natural Science Foundation of China (No.22409208) and the DICE Eureka Eurostars Project. Eurostars is part of the European Partnership on Innovative SMEs. The partnership is co-funded by the European Union through Horizon Europe. O.J.S. acknowledges funding from the Research Council of Finland through Project No. 357196. M.N. acknowledges funding from The Swedish Cultural Foundation in Finland. Philip Calado (University of Southampton) is acknowledged for fruitful discussions.